\documentclass[aps,prb,twocolumn,groupedaddress]{revtex4}
\usepackage{amsmath}
\usepackage{graphics}
\usepackage{graphicx}
\usepackage{epsfig}
\usepackage{amssymb}
\usepackage{amsmath}
\usepackage{amsfonts}
\begin{document}
\title{Universal behavior at discontinuous quantum phase transitions}
\author{Andr\'e S. \surname{Ferreira}}
\author{Mucio A. \surname{Continentino}}
\affiliation{Instituto de F\'{i}sica, Universidade Federal Fluminense \\
Campus da Praia Vermelha, Niter\'oi, 24210-340, RJ, Brazil}
\date{\today}
\begin{abstract}
Discontinuous quantum phase transitions besides their general
interest are clearly relevant to the study of heavy fermions and
magnetic transition metal compounds. Recent results show that in
many systems belonging to these classes of materials, the magnetic
transition changes from second order to first order as they
approach the quantum critical point (QCP). We investigate here
some mechanisms that may be responsible for this change.
Specifically the coupling of the order parameter to soft modes and
the competition between different types of order near the QCP. For
weak first order quantum phase transitions general results are
obtained. In particular we describe the thermodynamic behavior at
this transition when it is approached from finite temperatures.
This is the discontinuous equivalent of the non-Fermi liquid
trajectory close to a conventional QCP in a heavy fermion
material.
\end{abstract}
\maketitle

\section{Introduction}

Recently, the subject of discontinuous quantum phase transitions
has received much attention \cite{flouquet}. Besides the
theoretical interest in these transitions there are many new
experimental results showing that they occur in heavy fermions and
magnetic transition metal compounds \cite{first, Belitz}. In these
materials as one approaches a quantum critical point, the
transitions change their nature from second to first order
\cite{first}. There are many mechanisms which can drive a second
order transition into a first order one. For example, in
compressible magnets, pressure may produce this change
\cite{boccara}. In antiferromagnets \cite{boccara} and
superconductors \cite{paglione} this may be accomplished by an
external uniform magnetic field . Here we will be interested in
more subtle mechanisms which arise due to a coupling of the order
parameter to fluctuations~\cite{russos}. In this case the effects
are weaker being associated with what is generally known as {\em
weak first order transitions}. These mechanisms become more
relevant at very low temperatures where quantum fluctuations are
important.

The fluctuations we consider are of different types. In the first
case they are soft modes and as a paradigm of this case we study
the coupling of the superconductor order parameter to the
electromagnetic field. This has been investigated by Halperin,
Lubensky and Ma \cite{haluma} but, since we are interested in the
fully quantum mechanical case at zero temperature, our problem
becomes similar to that treated by Coleman and Weinberg in
particle physics \cite{Coleman}. Next, we consider the effect of
coupling the order parameter of a given phase to fluctuations of a
different phase which competes with the former in the same region
of the phase diagram. This problem is particularly relevant for
heavy fermion systems where an inhomogeneous state consisting of
antiferromagnetic and superconducting regions has been observed
\cite{kawasaki}. This coupling between competing order parameters
gives rise to new effects compared with the previous situation. In
this case not only the nature of the quantum transitions may
change but the ordered phase itself may become an inhomogeneous
mixture with different values for the order parameter.

We obtain general features of the behavior of a system approaching
a weak first order quantum phase transition (WFOQPT) from finite
temperatures. In spite that the correlation length does not
diverge as $T \rightarrow 0$, scaling does apply for these WFOQPT
and allows for a general description of this phenomenon.

\section{Coupling to soft modes: The Coleman-Weinberg
mechanism\label{CW-mechanism}}

\begin{figure}[h]
\begin{center}
\includegraphics{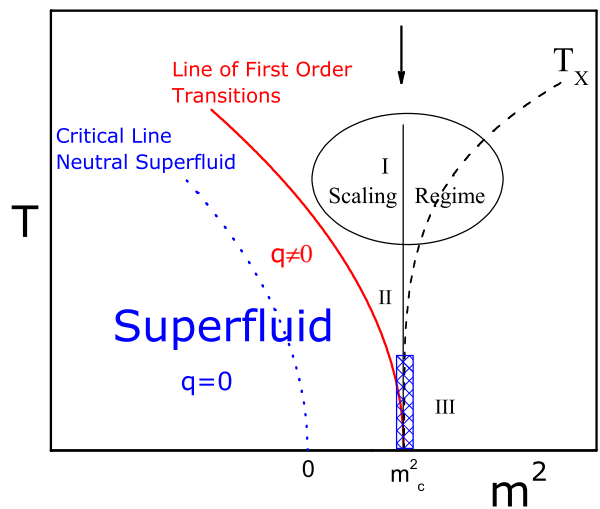}
\end{center}
\caption{Phase diagram of a charged superfluid coupled to photons. For
completeness we show also the critical line of the neutral superfluid. Along
the trajectory $m^{2}=m_{c}^{2}$ one can distinguish different regimes as
explained in the text.}%
\label{fig1}%
\end{figure}

A well known case of quantum fluctuations inducing a weak first
order transition is the Coleman -Weinberg mechanism
\cite{Coleman}. In the solid-state version, we consider a
superconductor represented by a complex field
$(\varphi_{1},\varphi_{2})$ coupled to the electromagnetic
field~\cite{MucioBook,PhysicaA}. The Lagrangian density of the
model is given by,
\begin{align}
L &  =-\frac{1}{4}(F_{\mu\nu})^{2}\!+\!\frac{1}{2}(\partial_{\mu}\varphi
_{1}\!+\!qA_{\mu}\varphi_{2})^{2}+\nonumber\label{SClagrangian}\\
&  +\frac{1}{2}(\partial_{\mu}\varphi_{2}-qA_{\mu}\varphi_{1})^{2}+\nonumber\\
&  -\frac{1}{2}m^{2}(\varphi_{1}^{2}+\varphi_{2}^{2})-\frac{\lambda}%
{4!}(\varphi_{1}^{2}+\varphi_{2}^{2})^{2}.
\end{align}
We are using $\hbar=c=1$ units and the indices $\mu,\nu$ run from
$0$ to $d=3$. In Eq.~(\ref{SClagrangian}) space and time are
isotropic and consequently the dynamic critical exponent $z=1$.
For a neutral superfluid $(q=0)$ the system decouples from the
electromagnetic field and has a continuous, zero temperature
superfluid-insulator transition at $m^{2}=0$ (see
Fig.~\ref{fig1}).

The effective potential method~\cite{Jona,ColemanBook,Jackiw}
yields the quantum corrections to the action given by the
Lagrangian density of Eq.~(\ref{SClagrangian}). At $T=0$ in the
one loop approximation, the effective potential close to the
transition ($m\approx0$) is given by~\cite{MucioBook}
\begin{equation}
V_{eff}=\frac{1}{2}m^{2}\varphi^{2}+\frac{\lambda}{4!}\varphi^{4}+\frac
{3q^{4}}{64\pi^{2}}\varphi^{4}\left[  \ln\left(
\frac{\varphi^{2}}{M^{2}}\right)  -\frac{25}{6}\right]
\label{eff1}
\end{equation}
where $M$ comes from renormalization and is completely
arbitrary~\cite{Coleman}. We can take the value of $M$ as the
minimum of the effective potential $\langle\varphi\rangle$. In
this case we have the equation, for $M=\langle\varphi\rangle$,
\begin{equation}
m^{2}+\frac{\lambda}{6}\langle\varphi\rangle^{2}-\frac{11q^{4}}{16\pi^{2}%
}\langle\varphi\rangle^{2}=0 \label{cond}%
\end{equation}
which yields
\begin{equation}
\langle\varphi\rangle^{2}=\frac{m^{2}}{\left(  \frac{\lambda}{6}-\frac
{11q^{4}}{16\pi^{2}}\right)  } \label{mean phi}%
\end{equation}
or alternatively for $\lambda$
\begin{equation}
\lambda=-\frac{6m^{2}}{\langle\varphi\rangle^{2}}+\frac{33q^{4}}{8\pi^{2}}.%
\end{equation}
We can use the equation above to remove $\lambda$ from the
effective potential Eq.~(\ref{eff1}) introducing another free
parameter, namely the non-trivial minimum of the effective
potential $\langle\varphi\rangle$. In this case, Eq.~(\ref{eff1})
reads
\begin{equation}
V_{eff}\!=\!\frac{1}{2}m^{2}\varphi^{2}\!-\!\frac{m^{2}}{4\langle
\varphi\rangle^{2}}\varphi^{4}\!+\!\frac{3q^{4}}{64\pi^{2}}\varphi^{4}\left[
\ln\left(  \frac{\varphi^{2}}{\langle\varphi\rangle^{2}}\right)
\!-\!\frac {1}{2}\right].  \label{eff2}
\end{equation}
The change of the parameters above known as dimensional
transmutation~\cite{Coleman} can only be done if the effective potential has a
non-trivial minimum $\langle\varphi\rangle\neq0$ as is clear from the
equations above. Since close to the transition $m\approx0$, we conclude from
Eqs.~(\ref{cond}) and~(\ref{mean phi}) that the condition for such a minimum
to exist is
\begin{equation}
\lambda\sim q^{4}. \label{lambda q}%
\end{equation}
Now let's define the coherence length
\begin{equation}
\xi=\sqrt{1/2|m^{2}|} \label{xi}%
\end{equation}
and the London penetration depth
\begin{equation}
\lambda_{L}=\sqrt{1/2q^{2}|\langle\varphi\rangle|^{2}} \label{lambda_L}%
\end{equation}
as is usually done for Ginzburg-Landau models. We can show that
the condition of Eq.~(\ref{lambda q}) is equivalent to
\begin{equation}
\lambda_{L}\ll\xi. \label{lambda_L xi}%
\end{equation}
This is the same condition for the occurrence of weak first order
transitions obtained in Ref.~\cite{haluma}. However, in general,
if $\lambda_{L}\sim\xi$, fluctuations of the order parameter can
destabilize the Coleman-Weinberg mechanism and the resulting
quantum transitions can be second order (see Ref.~\cite{Belitz}).

The condition~(\ref{lambda_L xi}) can be obtained from our results as follows.
The ratio $\lambda_{L}/\xi$, from Eq.~(\ref{xi}) and Eq.~(\ref{lambda_L}), is
\begin{equation}
\frac{\lambda_{L}}{\xi}=\sqrt{\frac{|m^{2}|}{q^{2}|\langle\varphi\rangle^{2}%
|}}.
\end{equation}
Substituting the result of Eq.~(\ref{mean phi}) we get
\begin{equation}
\frac{\lambda_{L}}{\xi}=\sqrt{\frac{1}{q^{2}}\left|  \frac{\lambda}{6}%
-\frac{11q^{4}}{16\pi^{2}}\right|  }.
\end{equation}
This ratio is independent of $m^{2}$ as expected and it is clear
that, if $\lambda\sim q^{4}$, we have $(\lambda_{L}/\xi) \ll1 $ or
$\lambda_{L} \ll\xi$ as expected. We will discuss more about this
condition later in section~\ref{RG}.

Let us consider that $\lambda\sim q^{4}$ (or equivalently $\lambda_{L}%
\ll\xi$) and study the minima of the effective potential,
Eq.~(\ref{eff2}). We find that at a critical value of the mass,
given by
\begin{equation}
m_{c}^{2}=\frac{3q^{4}}{32\pi^{2}}\langle\varphi\rangle^{2}%
\end{equation}
there is a first order transition to a superconducting state.
Notice that the transition in the neutral superfluid ($q=0$) is
continuous rather than first order and takes place at $m^{2}=0$.
Therefore, quantum fluctuations due to the coupling to soft modes,
the photons, lead to symmetry breaking in the normal region of the
phase diagram of the neutral superfluid.  This occurs close to the
quantum critical point of the chargeless system extending the
region where a symmetry broken phase appears in the
phase diagram. The transition occurs at a finite value of the mass $m_{c}%
^{2}$ and is first order. The shift of the critical mass depends
on a power of the coupling of the order parameter to the soft
modes, in the present case the charge of the Cooper pairs.

Introducing a parameter $g=m^{2}-m_{c}^{2}$ which measures the
distance to the transition, we can write the effective potential
at $T=0$ as,
\begin{equation}
V_{eff}= \frac{1}{4} \langle\varphi\rangle^{2}|m^{2}-m_{c}^{2}| \propto |g|^{2-\alpha}%
\end{equation}
with the exponent value $\alpha=1$ reflecting the fact the
transition is first order~\cite{PhysicaA}. The associated
\emph{latent heat} is $L_h =\frac{1}{4}
m_c^2\langle\varphi\rangle^{2}$. Spinodals points at $T=0$ can
also be calculated~\cite{PhysicaA}.

The finite temperature case can be studied replacing the frequency
integrations in the calculation of the effective potential by sums
over Matsubara frequencies~\cite{MucioBook}. The generalization of
the effective potential to finite temperatures close to the
transition can be written as~\cite{PhysicaA}
\begin{align}
V_{eff}(T) &  =\frac{1}{4}m^{2}\langle\varphi\rangle^{2}|g|\!\left[
1+\right.  \nonumber\\
&  \left.
+\frac{2}{\pi^{2}m^{2}\langle\varphi\rangle^{2}}\frac{T^{d+1}}
{|g|}I_{d}\!\left(  \frac{M(\varphi)}{T}\right)  \!\right]
\label{eff3}
\end{align}
where the integral $I$ is given by
\begin{equation} \label{I}
I_{d}(y)=\int_{0}^{\infty}dxx^{d-1}\ln{\left[
1-e^{-\sqrt{x^{2}+y^{2}} }\right]  }.
\end{equation}
and $M(\varphi)=m^{2}+q^{2}\varphi^{2}$. The function
$I_{3}(y)=I(y)$ can be obtained numerically integrating
Eq.~(\ref{I}).

The finite temperature phase diagram is shown in Fig.~\ref{fig1}. For
completeness we show the critical line of the neutral superfluid,
$T_{SF}=|m^{2}|^{\psi}$, which is governed by the shift exponent ${\psi}%
^{-1}=z/(d+z-2)=1/2$ in $d=3$ (see Ref.~\cite{MucioBook}). The new
line of first order transitions is given by
$T_{c}=\sqrt{m_{c}^{2}-|m^{2}|}$.

\section{Scaling at a weak first order quantum transition}

We will now consider the system \emph{sitting} at the new quantum
phase transition point, i.e., at $m^{2}=m_{c}^{2}$ and decrease
the temperature. This is the equivalent of the non-Fermi liquid
trajectory for ordinary quantum critical points in metallic
systems.

\begin{figure}[h]
\begin{center}
\includegraphics[height=5.8cm]{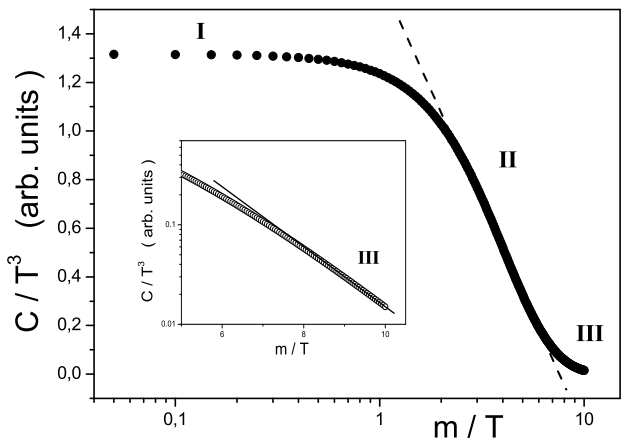}
\end{center}
\caption{The temperature dependent specific heat along the trajectory
$m^{2}=m_{c}^{2}$ of the zero temperature first order transition. The
different regimes are discussed in the text. Notice the dashed straight line
that indicates the \protect$\ln T$ behavior in regime II. }%
\label{fig2}%
\end{figure}

For high temperatures, $T \gg m_{c}$, which corresponds to the
regime I of Fig.~\ref{fig1} and Fig.~\ref{fig2}, the function
$I_{3}(y)$ saturates, $I_{3}(y < 0.12 )\approx- 2.16$. In this
case the effective potential,
\begin{align*}
V_{eff}(T) \approx\frac{1}{4}m^{2}\langle\varphi\rangle^{2}|g|\bigg\{ 1-\frac
{4.32}{\pi^{2}m^{2}\langle\varphi\rangle^{2}}\frac{T^{d+1}}{|g|}\bigg\}
\end{align*}
and can be cast in the scaling form,
\[
V_{eff}(T) \propto|g|^{2-\alpha} F \left[  \frac{T}{T_{\times}} \right]  .
\]
with $F(0)=$ $constant$. This scaling form is reminiscent of that
for the free energy close to a quantum critical point. In the
present case of a discontinuous zero temperature transition, the
critical exponent~\cite{PhysicaA} $\alpha=1$ and the
characteristic temperature is,
\[
T_{\times} \propto|g|^{\nu z} \propto|g|^{\frac{z}{d+z}} =
|g|^{\frac{1}{d+1}}=|g|^{\frac{1}{4}}
\]
with $\nu=1/(d+z)$~\cite{PhysicaA}. In this regime I or scaling
regime, along the line $m^{2}=m_{c}^{2}$ shown in Fig.~\ref{fig1},
the free energy density has therefore the scaling form $f(m=m_{c},
T) \propto T^{(d+z)/z} $ and the specific heat as shown in
Fig.~\ref{fig2} is given by,
\begin{equation}
C/T\big|_{(m=m_{c}, T)} \propto T^{\frac{d-z}{z}}.
\end{equation}
Then the thermodynamic behavior along the line $m^{2}=m_{c}^{2}$
in regime I ($T \gg m_{c}$) is \emph{the same} as when approaching
the quantum critical point of the neutral superfluid, along the
critical trajectory $m^{2}=0$. The system is unaware of the change
in the nature of the zero temperature transition and at such high
temperatures charge is irrelevant.

When further decreasing temperature along the line
$m^{2}=m_{c}^{2}$ there is an intermediate, non-universal regime
(regime II in Figs.~\ref{fig1} and \ref{fig2}). In the present
case for $m_{c} < T$ the specific heat $C/T^{d/z} \propto\ln T$ as
can be seen from the dashed straight line shown in the semi-log
plot of Fig.~\ref{fig2}.

Finally, at very low temperatures, for $T << m_{c}$ and
$m^{2}=m_{c}^{2}$, i.e., in regime III of Fig.~\ref{fig2}, the
specific heat vanishes exponentially with temperature, $C/T^{d/z}
\propto\exp(-m_{c}/T)$. The gap for thermal excitations is given
by the shift $m_{c}$ of the quantum phase transition. The
correlation length which grows along the line $m^{2}=m_{c}^{2}$
with decreasing temperature
reaches saturation in regime III at a value $\xi_{S} \approx m_{c}^{-1}%
$. Then, we can understand the exponential dependence of the
specific heat as due to gapped excitations inside superconducting
bubbles of finite size $L \thicksim\xi$. The gap between the
states in the bubbles is $\Delta\sim L^{-z} \sim\xi^{-z} \sim
m_{c}$ as we have found previously.

Although the results above have been obtained for a particular
model, the behavior in the scaling regime I and III should be
universal and characteristic of any weak first order quantum
transition. Notice that the relevant critical exponents which
determine the scaling behavior in particular in regime I are those
associated with the QCP of the {\em uncoupled system} which in the
present case is the neutral superfluid.

\section{Coupling between order parameters}

\subsection{Quantum first order transitions in superconducting heavy-fermions\label{SCandAF}}

Weak first order transitions and spontaneous symmetry breaking
near quantum phase transitions can also occur due to the coupling
of an order parameter to nearly critical fluctuations of another
phase. This provides a new mechanism for WFOQPT besides the
coupling to  gapless excitations as treated above and in
Ref.~\cite{Belitz} for example (see also
Refs.~\cite{note,berker}).

In this section we study the coupling, at $T=0$, of an order
parameter to fluctuations of a competing order. This type of
coupling becomes important when two different phases are in
competition on the same region of the phase diagram. In this case
for some values of the microscopic interactions which are close,
there are alternative ground states which interfere with one
another. This is the situation, for example, of several
heavy-fermions~\cite{Pagliuso} and of high-T$_c$
superconductors~\cite{Kastner} where in general the dispute is
among superconductivity and antiferromagnetism. We obtain the
effects on a given phase, of the coupling of its order parameter
to fluctuations of the competing order, in the form of quantum
corrections. We show that these can produce spontaneous symmetry
breaking and change the nature of the transition even if these
fluctuations are non critical. Again, due to the weak first order
character of the new phase transition we expect to find scaling
behavior as described previously.

\begin{figure}[h]
\begin{center}
\includegraphics[height=3.5cm]{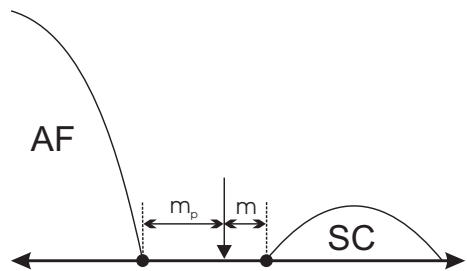}
\end{center}
\caption{Normal phase separating SC and AF phases.}%
\label{diagram}
\end{figure}

We consider the case of  a heavy-fermion~(HF) system with a
superconducting phase that is close to an antiferromagnetic
instability at $T=0$~\cite{PRB} as in Fig.~\ref{diagram}. As
before, we first describe the superconducting phase by the
simplified Lorentz invariant free action as in
Eq.~(\ref{SClagrangian}) but now coupled to an AF order parameter
instead of the electromagnetic field. In many cases however the
effects of dissipation should be taken into account. Then, later
on we use another free action for the superconducting field. This
different quadratic form is associated with a dynamic exponent
$z=2$, instead of $z=1$ of the Lorentz invariant
case~\cite{Ramazashvili,Kirkpatrick}.

Our model is of the Ginzburg-Landau type and contains three real
fields. Two fields, $\phi_{1}$ and $\phi_{2}$, correspond to the
two components of the superconductor order parameter. The other
field $\phi_{3}$, for simplicity represents a one component
antiferromagnetic order parameter. The results are immediately
generalized to a three component AF vector field, with the unique
consequence of changing some numerical factors~\cite{PRB}. The
free functional associated with the magnetic part represented by
the field $\phi_{3}$, the sub-lattice magnetization, takes into
account the dissipative nature of the paramagnons near the
magnetic phase transition~\cite{Hertz} in the metal and yields the
propagator
\begin{equation}
\label{propagatorAF}D_{0}(\omega,\boldsymbol{q})=\frac{i}{i|\omega|\tau
-q^{2}-m_{p}^{2}}
\end{equation}
where $\tau$ is a characteristic relaxation time and $m_{p}^{2}$
is related to a local Coulomb repulsion $U$ and the density of
states at the Fermi level $N(E_{F})$ by
\begin{equation}
m_{p}^{2}=1-UN(E_{F}).
\end{equation}

The part of the action associated with the potential is given by
\begin{align}
\label{pot}V_{cl}(\phi_{1},\phi_{2},\phi_{3})=\frac{1}{2}m^{2}(\phi_{1}
^{2}+\phi_{2}^{2})+\frac{1}{2}m_{p}^{2}\phi_{3}^{2} +\nonumber\\
+ V_{s}(\phi_{1},\phi_{2}) + V_{p}(\phi_{3}) + V_{i}(\phi_{1},\phi_{2}
,\phi_{3}),
\end{align}
where the self-interaction of the superconductor field is
\begin{equation}
V_{s}(\phi_{1},\phi_{2})=\frac{\lambda}{4!}(\phi_{1}^{2} + \phi_{2}^{2})^{2},
\end{equation}
and that of the antiferromagnet is given by
\begin{equation}
V_{p}(\phi_{3})=\frac{g}{4!}\phi_{3}^{4}.
\end{equation}
Finally, the last term is the (minimum) interaction between the relevant
fields,
\begin{equation}
\label{coupling}V_{i}(\phi_{1},\phi_{2},\phi_{3})=u(\phi_{1}^{2}+\phi_{2}%
^{2})\phi_{3}^{2}.
\end{equation}
This term is the first allowed by symmetry on a series expansion
of the interaction. Notice that for $u>0$, which is the case here,
superconductivity and antiferromagnetism are in competition and
this term does not break any symmetry of the original model.
However, including quantum fluctuations we show that spontaneous
symmetry breaking can occur in the normal phase separating the SC
and AF phases. This case is represented schematically in
Fig.~\ref{diagram} and is discussed in the subsequent sections.

\subsection{Superconducting transition\label{SCtransition}}

We first consider the metal-superconductor transition in HF when
the superconducting field is coupled, as in Eq.~(\ref{coupling}),
to AF fluctuations described by the propagator of
Eq.~(\ref{propagatorAF})(Fig.~\ref{diagram}). Detailed calculation
of the effective potential have already been presented~\cite{PRB}.
The general result is given by,
\begin{equation}
\label{potSC}V_{ef}(\phi_{c})
\approx\frac{1}{2}M^{2}\phi_{1c}^{2}+a
m_{p}^{2}\phi_{1c}^{2}|\phi_{1c}|+\frac{\tilde{\lambda}}{4!}\phi_{1c}^{4}+\mathcal{O}(\phi
^{5}).
\end{equation}
In Eq.~(\ref{potSC}), $\tilde{\lambda}$ is a renormalized coupling
but of the same order of the bare coupling $\lambda$. The new
coupling $a$ introduced by fluctuations can produce a symmetry
breaking in the normal state extending the SC region in the phase
diagram at $T=0$. The same mechanism turns the transition to weak
first order with a small latent heat\cite{S.Stat.Comm.}. Using the
same argument as in the preceding section we can introduce again
parameters associated with $\xi$ (coherence length) and
$\lambda_{L}$ (London penetration depth) in this case given by
\begin{align}
\xi= \sqrt{1/2|m^{2}|}\\
\lambda_{L} = \sqrt{1/u|\langle\phi\rangle^{2}|}.
\label{lambda_L2}
\end{align}
It's difficult to carry out the dimensional transmutation
analytically on the result of Eq.~(\ref{potSC}) since the
equations of minima are hard to solve. However, we can check
numerically that the condition for the existence of minima away
from the origin is equivalent to $\lambda_{L}\ll\xi$ as in the
previous case. It's also interesting to notice that the first
order transition is produced by the cubic term in
Eq.~(\ref{potSC}) and this term is proportional to the {\it
magnetic mass} $m_{p}$. If magnetic fluctuations were critical,
i.e.,  $m_{p}=0$, the only effect of the coupling would appear as
a term proportional to $\phi^{5}$. This term is usually neglected
since its power is higher than those initially considered in the
classical potential and usually insufficient to create new minima
around the origin. Therefore, if AF fluctuations were critical the
effects of the quantum corrections in the transition could be
neglected.

\subsection{Antiferromagnetic transition}

In order to study the magnetic transition, as mentioned previously
we consider two kinds of quadratic forms associated with the free
superconducting fields. The first is the  usual Lorentz invariant
free action that we have discussed before in Sec.~\ref{SCandAF}.
Next we work with another free action which takes into account
dissipation and is associated with a $z=2$ dynamics.

Let's first study the result for the Lorentz invariant propagator. Close to
the AF transition we have~\cite{PRB}
\begin{align}
\label{potAF}V_{ef}(\phi_{3c})  &  \approx\frac{1}{2}M_{p}^{2}\phi_{3c}^{2}+
+\frac{\tilde{g}}{4!}\phi_{3c}^{4}+\tilde{u}^{2}\phi_{3c}^{4}\ln\left(
\frac{\phi_{3c}^{2}}{\langle\phi_{3}\rangle^{2}}\right) \nonumber\\
&  +\tilde{u}m^{2}\phi_{3c}^{2}\ln{\left(
\frac{\phi_{3c}^{2}}{\langle \phi_{3}\rangle^{2}}\right)  }
\end{align}

We notice directly from Eq.~(\ref{potAF}) that if the
superconductor order parameter fluctuations were critical we would
obtain the same result of Eq.~(\ref{eff2}) since in this case the
parameter $m^{2}=0$. However, SC fluctuations are not critical but
close to criticality and the last term of Eq.~(\ref{potAF}) may
become important. Of course, its relevance depends on the strength
of the renormalized coupling $\tilde{u}$ and the results
considering this term leads to new and interesting changes in the
ground state. We obtain, besides the two finite minima of the
Coleman-Weinberg potential of Eq.~(\ref{eff2}),  two extra minima
very close to the origin~\cite{PRB}. The normal state associated
with these minima has a small moment since the value of the order
parameter is related to the sub-lattice magnetization. An
additional first order transition occurs when the other two minima
away from the origin become the stable ones as the system moves
away from the superconductor instability. This transition is from
a small moment AF (SMAF) to a large moment AF (LMAF) and occurs
even before the continuous mean field transition. When we move
away from the magnetic transition the strength of this new term
decreases with the value of $m^{2}$ and the two new minima move to
the origin producing a normal state with vanishing magnetization
again. It's important to note here that the SMAF is obtained
because the magnetic order parameter couples to fluctuations which
are close to criticality but not critical. Critical fluctuations
yield the same results of section~\ref{CW-mechanism}.

Now, for many cases of interest, SC fluctuations are better
described by a dissipative propagator associated with a $z=2$
dynamics\cite{Ramazashvili,Kirkpatrick} similar to
Eq.~(\ref{propagatorAF}). This is useful to account for pair
breaking interactions as magnetic impurities that can destroy
superconductivity~\cite{Sigrist, Mineev}. It is given by,
\begin{equation}
G_{0}(\omega,\boldsymbol{q})=\frac{i}{i|\omega|\tau^{\prime}-q^{2}-m^{2}}.
\end{equation}
The parameter $m^{2}$ is still related with the classical distance
from the phase transition and we have a relaxation time
$\tau^{\prime}$. In general we also have a non-dissipative term
(as in the case studied above) but this term is neglected since
the linear term is the most important in the low frequency region.
Calculation of the effective potential is very similar to the
previous cases and the result has the form
\begin{equation} \label{Veffz=2}
V_{eff}=\frac{1}{2}M_{p}^{2}\phi_3^{2}+\frac{1}{4!}\tilde{g}\phi_3^{4}+\frac
{1}{15\pi^{2}}(2u\phi_3^{2}+m^{2})^{5/2}
\end{equation}
where the renormalized mass is
\begin{equation}
M_{p}=m_{p}^{2}-\frac{2}{3\pi^{2}}m^{3}u-\frac{1}{12\pi^{2}}m_{p}^{3}g
\end{equation}
and the renormalized coupling
\begin{equation}
\tilde{g}=g-\frac{12}{\pi^{2}}mu^{2}-\frac{3}{8\pi^{2}}m_{p}g^{2}.
\end{equation}
Quantum corrections can once again produce a weak first order
transition. We note that in this case the renormalization of the
potential obtained is easier since there are no logarithmic terms
to renormalize~\cite{PRB}. Anyway, the result is very similar to
that presented in section~\ref{SCtransition} since we have
equivalent propagators for the fluctuations. In general the form
of the quantum correction depends not only on the effective
dimension of the uncoupled QCP but also on the form and specially
the dynamics of the  soft modes or competing fluctuations.

The transformation $(\phi_3')^2=2u\phi_3^{2}+m^{2}$ makes the
potential of Eq.~(\ref{Veffz=2}) simpler and the analysis of its
extrema shows that the transition can be first order depending on
the coupling values. The appearance of SMAF phases is not possible
in this case. We note, however, that the couplings are dependent
on the renormalization parameter $M$ defined in
Sec.~\ref{CW-mechanism} and a renormalization group approach is
essential to make the results more reliable. In the next section
we illustrate this with an analysis of the Coleman-Weinberg case.

\section{One-loop effective potentials and renormalization group\label{RG}}

In the previous sections we have obtained weak quantum first order
transitions from minima generated balancing terms in the effective
potential. This is possible when we compare terms of the same
order and therefore the minima depend on the values of the
couplings. Considering, for example, the superconductor coupled to
the electromagnetic field, the terms we have to balance are
proportional to $\lambda$ and $q^{4}$. New minima away from the
origin are produced if
\begin{equation}
\label{conRG}\lambda\sim q^{4}
\end{equation}
and this condition also leads to a relation for the London
penetration depth and the coherence length $\lambda_{L}\ll\xi$. In
this section we show how we can use renormalization group
arguments to generalize the results for any small $\lambda$ and
$q$ and get rid of the condition~(\ref{conRG}).

When deriving Eq.~(\ref{eff1}) we have introduced a new parameter
$M$, associated with a renormalization mass and which was
completely arbitrary. The results for the model must be the same
for any chosen value of $M$. Now, if we prove that a variation in
the value of $M$ produces a correspondent variation in the
coupling constant $\lambda$, it's always possible to satisfy the
condition~(\ref{conRG}) choosing a suitable value of $M$. To prove
that, we have to consider Eq.~(\ref{eff1}). We want to rewrite
this equation with a value $M^{\prime}$, such that,
\begin{align}
V_{eff}  &  =\frac{1}{2}m^{2}\varphi^{2}+\frac{\lambda}{4!}\varphi^{4}+
\frac{3q^{4}}{64\pi^{2}}\varphi^{4}\left[  \ln\left(  \frac{\varphi^{2}%
}{{M^{\prime}}^{2}}\frac{{M^{\prime}}^{2}}{M^{2}}\right)  -\frac{25}{6}\right]
\nonumber\\
V_{eff}  &  =\frac{1}{2}m^{2}\varphi^{2}+\frac{\lambda}{4!}\varphi^{4}%
+\frac{3q^{4}}{64\pi^{2}}\varphi^{4}\ln\frac{{M^{\prime}}^{2}}{M^{2}
}+\nonumber\\
&  \frac{3q^{4}}{64\pi^{2}}\varphi^{4}\left[  \ln\left(  \frac{\varphi^{2}%
}{{M^{\prime}}^{2}}\right)  -\frac{25}{6}\right] \nonumber
\end{align}
which is equivalent to Eq.~(\ref{eff1}) with the reparametrization
\begin{equation}
\frac{\lambda^{\prime}}{4!}=\frac{\lambda}{4!}+\frac{3q^{4}}{64\pi^{2}}%
\ln\frac{{M^{\prime}}^{2}}{M^{2}}.
\end{equation}
In this sense the effective potential is always given by the same equation
with a suitable parametrization of the coupling.

This argument can be formally stated in an renormalization group
approach and we refer to the correspondent section of
Ref.~\cite{Coleman}. As a result, a weak first order transition
occurs for any small values of $\lambda$ and $q$, since we can
appropriately choose the renormalization mass. The only
restriction comes from perturbation theory which requires small
couplings.

For the case of coupling between AF and SC order parameters we can
also construct a renormalization group. Within this approach we
expect to find all the conditions for the occurrence of weak first
order transitions. Detailed calculations and results will be
published elsewhere.

\section{Conclusions}

We have studied the effects on a phase transition of coupling the
order parameter of this phase to soft modes or to fluctuations of
a competing phase. Since  these effects become more important at
low temperatures we have mostly considered the zero temperature
case. We have obtained the effects of quantum corrections due to
this coupling on the quantum critical point associated with the
relevant order parameter. These corrections were calculated using
the effective potential method. They can produce radical
modifications on the nature of the original phase transition. They
can change the nature of the transition, from continuous to
discontinuous and also affect the ordered phase itself giving rise
to an inhomogeneous phase with two values of the order parameter.

Whenever the fluctuations  change the nature of the transition to
weak first order, we can study them using a scaling approach. We
have obtained explicit results for one particular case but that
should hold in general. As the system approaches the discontinuous
transition from non-zero temperature, we can distinguish three
different regimes. In the first, the high temperature regime, the
scaling behavior is the same as approaching the quantum critical
point of the uncoupled system, i.e., in the absence of additional
fluctuations or soft modes. Further decreasing the temperature
there is a non-universal regime which may depend on the particular
dynamics of the original system and of the nature of the
additional modes. Finally at the lowest temperatures there is a
regime which is characteristic of the first order nature of the
zero temperature transition. At such low temperatures the
correlation length saturates at a finite value. The thermodynamic
properties have a contribution which is thermally activated. This
can be traced to excitations which are gapped due to finite size
effects as they occur in finite bubbles of the incipient phase.
The correctness of this interpretation is clear from the fact that
the gap for excitations is inversely related to the size of the
bubbles which in turn is of the order of the correlation length.
Notice also that the former is related to the shift in the zero
temperature phase transition with respect to that of the uncoupled
system.

The relevant exponents which determine the scaling behavior near
the weak first order transition are those associated with the
quantum critical point of the uncoupled system, i.e., of the
system in the absence of competing fluctuations or soft modes. In
particular this is the case of the dynamic exponent. The scaling
properties of the new transition are related to the existence of
an underlying second order instability which has become
discontinuous due to the effects of soft modes or of fluctuations
of a competing order parameter.

\begin{acknowledgments}
The authors would like to thank the Brazilian Agencies, FAPERJ and
CNPq for partial financial support. MAC acknowledges Dr. V. Mineev
for enlightening discussions.
\end{acknowledgments}


\begin{thebibliography}{99}
\bibitem{note} The concept of generic scale invariance (GSI) has been emphasized
recently with respect to systems with continuous broken symmetry
\cite{Belitz}. In the simplest case of a Heisenberg ferromagnet in
an external magnetic field ($d > d_{c}=2$) GSI is related to the
strong coupling fixed point at $(T/J)=0$, $h=0$ which
governs the symmetry broken phase. Close to this attractor, $J^{\prime}%
=b^{d-2}J$ and $h^{\prime}=b^{d} h$. The free energy density
scales as, $ f \propto b^{-d} G[(h/J)b^{2}]$. Then $f \propto
h^{d/2}$, the magnetization $m=\partial f/\partial h \propto
h^{\frac{d-2}{2}}$ and the susceptibility $\chi\propto
h^{\frac{d-4}{2}}$. An additional contribution arises at $T/J=0$,
$h=0$,  namely $ f= b^{-d}g[b^{d}h] $. This yields $f \propto|h|$
and $m = \pm m_0$ for $h \rightarrow \pm 0$ \cite{berker}.
\end{thebibliography}
\end{document}